\documentclass{aa}
\def\simle{\lower 2pt \hbox {$\buildrel < \over {\scriptstyle \sim }$}}
\def\simge{\lower 2pt \hbox {$\buildrel > \over {\scriptstyle \sim }$}}
\begin{document}
\thesaurus{
            13.07.1;      
            02.19.1;      
            09.10.1;      
            02.18.5}      

\title{A jet-disk symbiosis model for Gamma Ray Bursts: SS 433 the next?}
\author{G. Pugliese \and H. Falcke \and P. L. Biermann}
\institute{Max-Planck-Institut f\"{u}r Radioastronomie, Auf dem
H\"{u}gel 69, D-53121 Bonn, Germany}
\offprints{G. Pugliese, pugliese@mpifr-bonn.mpg.de}
\date{Received December 17 1998 , Accepted March 5 1999}
 \maketitle\markboth{Pugliese et al.: A jet-disk symbiosis model for Gamma
 Ray Bursts: SS 433 the next?}
{Pugliese et al.: A jet-disk symbiosis model for Gamma
 Ray Bursts}

\begin{abstract}

We consider a jet-disk symbiosis model to explain Gam\-ma Ray Bursts
and their afterglows. It is proposed that GRBs are created inside a
pre-existing jet from a neutron star which collapses to a black hole
due to massive accretion.  In our model we assume that the initial
energy due to this transition is all deposited in the jet by magnetic
fields, using fully well explored concepts from jets and disks in 
active galactic nuclei and compact active stars in binary systems. The 
observed emission is then due to an ultrarelativistic shock wave 
propagating along the jet. We show that a good agreement between model
predictions and observational data can be obtained for systems with 
accretion rates as high as in the Galactic jet source SS433. 
Specifically, we are able to reproduce the typical observed afterglow 
emission flux and its spectrum as a function of time. 

\end{abstract}

\section{Introduction}

Gamma Ray Bursts (GRBs) have been a mystery for almost 30 years.
Recently, thanks to the Italian-Dutch satellite BeppoSax, it has been
possible to detect for the first time their X-ray, optical, and radio
afterglows. Many articles have been published to report the main
characteristics of GRBs, (e.g., Guarnieri \& al. 1997; Piro \&
al. 1998; Frail \& al. 1997; Metzger \& al. 1997; Gorosabel \&
al. 1998; Kulkarni \& al. 1998).  By now we know that GRBs are
isotropically distributed over the sky, at least three of them are
cosmological, they show a time variability in the $\gamma$-emission
of the order of milliseconds, and long complex bursts. Many, but not all 
of them have an X-ray afterglow, and it seems that for most of them 
a host galaxy can be found.

Modern theoretical attempts to interpret the data are based on ideas
by M{\'e}sz{\'a}ros and Rees (1993), M{\'e}sz{\'a}ros (1994),
Panaitescu and M{\'e}sz{\'a}ros (1998), as well as Paczy{\'n}ski
(1986), and Paczy{\'n}ski and Rhoads (1993). In these models a
relativistic shock is caused by a relativistic fireball in a
pre-existing gas, such as the interstellar medium or a stellar wind,
producing and accelerating electrons/po\-si\-trons to very high
energies, which produce the gamma-emission and the various
afterglows observed. The low level of associated radiation at other
wavelengths limits the baryonic load of the emitting regions to very
low amounts, and constrains the scale of the emitting region to
lengths much larger than a neutron star.

One serious question is whether the overall energetics of the
fireball---assumed to be isotropic---are reasonable {(Sari and Piran
1997; Dar 1997)} or actually exceed the level given by any conceivable
model of neutron star mergers or other stellar collapses.  Another
question is whether a fireball model that uses external or internal
shock waves can solve the baryonic mass load problem to explain the
$\gamma$-emission, starting with an initial energy of $10^{51}$ erg in
the spherical shell rest frame.

Many authors suppose the validity of the fireball model and provide
evidence for its agreement with the observations (Waxman 1997a, Waxman
1997b, Vietri 1997), but the observational results for GRB971214,
requiring an initial energy $E \simeq 10^{53}$ erg (if the emission is
isotropic), pose a serious challenge to existing models.

Here we propose an anisotropic model where the $\gamma$-ray emission
and the afterglow is produced inside a preexisting jet and calculate
the temporal evolution of the corresponding flux.

\section{GRB jet model}

In our model GRBs develop in a pre-existing jet. We consider a binary
system formed by a neutron star and an O/B/WR companion in which the
energy of the GRB is due to the accretion-induced collapse of the
neutron star to a black hole.

The high speed in the energy flow in AGN-jets is generally believed to be
initiated by strong magnetic field coupling to the accretion disk of the black
hole (e.g. Falcke $\&$ Biermann 1995, Romanova \& Lovelace 1997, Romanova et 
al. 1997). We use the same mechanism here. We assume that in this transition a 
large amount of energy is anisotropically released as Poynting flux along the 
polar axis. One may think of this process as a violent and rapid twist, such 
as occasionally debated as a possible cause for supernova explosions (Kardashev 
1970, Bisnovatyi-Kogan 1970, Le Blanc \& Wilson 1970, and more recently 
Biermann 1993). This energy release naturally initiates an ultrarelativistic 
shock wave in the pre-existing jet. The emission microphysics are as in 
existing fireball models.

To fix the jet parameters, we use the basic ideas of the jet-disk
symbiosis model by Falcke $\&$ Biermann (1995, 1999).  In this model
the direction of the magnetic field is mostly perpendicular to the 
axis of the jet and the values of the total particle number $n_{\rm j}$
(relativistic electrons + thermal protons) and magnetic field $B_{\rm j}$
are calculated using the equipartition between magnetic field energy
density and kinetic plasma energy density in the umperturbed jet. Both, 
$B_{\rm j}$ and $n_{\rm j}$, are a function of the jet ejection rate $\dot 
M_{\rm jet} \simeq 0.05 \dot M_{\rm disk} \simeq ({10^{-5} {M_{\rm 
{\odot}}/{\rm{yr}}}}) {\dot M_{-5 \rm j}}$, the equipartition parameter 
$\epsilon$ and the bulk velocity of the jet $\beta_{\rm j} \gamma_{\rm j} = 
0.3 v_{0.3}$, where $\gamma_{\rm j}$ is the bulk Lorentz factor of the jet 
prior to the burst.  The values of the particle number density and the 
magnetic field in the unperturbed jet are then given as

\begin{equation}
n_{\rm j}(z_{\rm j}) \simeq 80.34 \; (\dot M_{-5 \rm j} v_{0.3}^{-1}) 
\theta_{-1 \rm j}^{-2} z^{-2}_{17,{\rm j}} \quad {\rm{cm^{-3}}}
\end{equation}

\begin{eqnarray}
B_{\rm j}(z_{\rm j}) 
& = & 1.05 \; (\dot M_{-5 \rm j}^{1/2} v_{0.3}^{-1/2}) 
     (\epsilon^{-1/2} \theta_{-1 \rm j}^{-1} 
     \gamma_{{\rm m},2}^{1/2}) \times \nonumber \\
&   & z^{-1}_{17, {\rm j}} \quad {\rm{Gauss}}
\end{eqnarray}
where $\gamma_{{\rm m},2}$ is the minimum electron Lorentz factor 
$\gamma_{\rm em} = 100 \gamma_{{\rm m},2}$ of the relativistic electrons 
(assumed to be in a power law distribution), $\theta_{-1 \rm j} = 
\theta_{\rm j}/(0.1 {\rm{rad}})$ is the opening angle of the jet, and 
$z_{\rm j} = 10^{17} z_{17, {\rm j}} \, \rm cm$ is the distance along the jet. 
$z$ is the redshift.

The advantages of using the propagation of an ultrarelativistic shock
wave inside a jet can be summarized in the following two points

\begin{description}

\item [$\bullet$] The initiation of the shock by magnetic fields does
not necessarily involve considerable matter. It implies a low amount 
of baryonic matter, of order $5 \times 10^{-8} \; M_{\rm {\odot}}$, in the jet.

\item [$\bullet$] The initial amount of energy deposited in the jet is of the 
order $E_{51} = E/(10^{51} {\rm{erg}} )$, and we can obtain an apparent 
isotropic energy as high as $E_{\rm {app}} = 10^{53.3} E_{51} 
\theta^{-2}_{-1 \rm j}$ ${\rm{erg}}$. This fits well the requirements 
of GRB971214.

\end{description}

The evolution of the shock Lorentz factor $\gamma_{\rm sh}$ with distance
$z_{\rm j}$ along the jet axis can be obtained using the conservation
of the unperturbed jet gas energy in the shock rest frame, $E/2 = \pi
{\theta_{\rm j}}^2 z_{\rm j}^3 \gamma_{\rm sh}^2 m_{\rm p} c^2 n_{\rm j}$, 
that is

\begin{equation}
\gamma_{\rm sh}(z_{\rm j}) \approx 11.48 \; (E^{1/2}_{51} \dot 
M_{-5 \rm j}^{-1/2} v_{0.3}^{1/2}) z^{-1/2}_{17, {\rm j}}.
\end{equation}

The characteristic time scale to see the emission across the entire region
when the shock has reached a distance $z_{\rm j}$ along the axis of the jet
is given by $t^{\rm{(ob)}} = z_{\rm j}/(2 \gamma_{\rm sh}^2c)$. Substituting 
this into the formula for $\gamma_{\rm sh}$, we get the time evolution of the 
ultrarelativistic shock front

\begin{equation}
z_{\rm j}(t) \approx 2.81 \times 10^{17} (E^{1/2}_{51} \dot M_{-5 \rm j}^{-1/2}
v_{0.3}^{1/2}) t_5^{1/2} \quad \rm{cm}
\end{equation}

\begin{equation}
\gamma_{\rm sh}(t) \approx  6.84 \; (E^{1/4}_{51} \dot M_{-5 \rm j}^{-1/4}
v_{0.3}^{1/4}) t_5^{-1/4},
\end{equation}
where $t_5 = t/(10^5 \rm{s})$.

A change in the emission properties occurs when the opening angle of our jet 
passes the relation $1/ \gamma_{\rm sh} = {\theta_{\rm j}}$. Using the
formula of the bulk Lorentz factor as a function of time, we can see at what 
time $t^{\rm{\star}}$ this relation holds: $t^{\rm{\star}} \approx 2.20 \times 
10^4 (E_{51} \dot M_{-5 \rm j} v_{0.3}) \theta_{-1 \rm j}^4$ s. Hence, after 
about 6 hours this limit is reached.  Prior to this, the observed emission 
is limited by the Lorentz boost to a conical section of the shock front of 
angle $1/\gamma_{\rm sh}$. For the following calculation of the afterglow 
emission several hours after the burst we can therefore consider $1/ 
\gamma_{\rm sh} > {\theta_{\rm j}}$.

In the flow behind a steady shock front, relativistic particles are
usually accelerated and magnetic fields can be amplified. After the 
shock, $B_{\rm {\Vert}}^{\prime} \simeq B_{\rm {\Vert}}$ and 
$B_{\rm {\bot}}^{\prime} \simeq 4 \gamma_{\rm sh} B_{\rm {\bot}}$, 
relative to the jet axis. The resulting magnetic field in the shocked 
plasma will have a strength of roughly $B \approx {(16 \gamma_{\rm sh}^2 + 
1)}^{1/2} B_{\rm {\bot}}$ and the shock wave compresses the magnetic 
field component perpendicular to the jet axis. The jump conditions for the 
density particle number give $n^{\rm{(sf)}} \simeq 4 \gamma_{\rm sh} 
n_{\rm j}$, (De Hoffmann and Teller 1950, Marscher and Gear 1985). We use here 
the approximation for an ultrarelativistic shock so that $n_2/n_1 = 4 
\gamma_{12} \simeq 4 \gamma_{\rm sh}$, where 1 and 2 are related to the zone 
before and after the shock. This allows a straight and simple limit to 
nonrelativistic shocks. In our model, this corresponds to

\begin{eqnarray}
n_{\rm j}^{' \rm {(sf)}}(t) 
& = & 2.78 \times 10^2 (E_{51}^{-3/4} \dot M_{-5 \rm j}^{7/4}
      v_{0.3}^{-7/4}) \times \nonumber \\
&   & \theta_{-1 \rm j}^{-2} t_5^{-5/4} \quad \rm{cm^{-3}}
\end{eqnarray}
\begin{eqnarray}
B_{\rm j}^{' \rm {(sf)}}(t) 
& = & 10.24 \; (E_{51}^{-1/4} \dot M_{-5 \rm j}^{3/4}
      v_{0.3}^{-3/4}) \times \nonumber \\
&   & (\epsilon^{-1/2} \theta_{-1 \rm j}^{-1} 
      \gamma_{{\rm m},2}^{1/2}) t_5^{-3/4} \quad \rm{Gauss}
\end{eqnarray}
where (sf) shows the quantities in the shock frame and (ob) the
corresponding values in the observer frame.

In our jet we consider an electron power law distribution $N(\gamma_{\rm e})
d \gamma_{\rm e} = C_{\rm e} \gamma^{\rm {-p}}_{\rm e} d \gamma_{\rm e}$ where 
$\gamma_{\rm em} < \gamma_{\rm e} < \gamma_{\rm max}$, with a cutoff at a 
constant minimum electron Lorentz factor $\gamma_{\rm em} \simeq 100$. In 
fact, if p-p collisions inject a population of electrons in the unperturbed 
jet, then one would expect that in the unperturbed jet the electron Lorentz 
factor $\gamma_{\rm e}$ goes from $\gamma_{\rm em} \simeq 100 \simeq m_{\rm 
{\pi}}/m_{\rm e}$ up to some large value (Falcke \& Biermann 1995).

The equation for $C_{\rm e}$ has been obtained considering that in the shock
frame the powerlaw electron energy density is taken as a fraction
$\delta \leq 1$ of the pre-existing relativistic electron energy density
and using the value p=2 

\begin{eqnarray}
C_{\rm e}^{\rm{(sf)}}(t) 
& = & 1.91 \times 10^4 \delta (E_{51}^{-1/2} \dot M_{-5 \rm j}^{3/2} 
      v_{0.3}^{-3/2}) \times \nonumber \\
&   & (\theta_{-1 \rm j}^{-2} \gamma_{{\rm m},2}) t_5^{-3/2}
      \quad \rm{cm^{-3}}
\end{eqnarray}
If we were to relate to the proton energy density in the shock, 
$\delta \le 20 \simeq m_{\rm p}/(m_{\rm e} \gamma_{\rm em})$.

To calculate the boosting factor of the transition from the shock frame
to observer frame, we assume the angle between the jet-axis and the line
of sight to the observer is $\theta_{\rm {obs}} < 1/ \gamma_{\rm sh}$ some time
during the $\gamma$-ray burst. The critical synchrotron frequency after 
the shock front in the observer frame is given by $\nu_{\rm m} \approx {2 \over 
{1 + z}} \gamma_{\rm sh}^3 \gamma^2_{\rm em} {{e B} \over m_{\rm e} c}$, 
where we consider the minimum electron Lorentz factor evolves in phase space 
with the bulk Lorentz factor $\gamma_{\rm em} \times \gamma_{\rm sh}$ : 

\begin{equation}
\nu_{\rm m} (t) \approx {{2.61 \times 10^{14}} \over {1 + z}} E_{51}^{1/2} 
(\epsilon^{-1/2} \theta_{-1 \rm j}^{-1} \gamma_{{\rm m},2}^{5/2}) t_5^{-3/2} 
\quad \rm{Hz}
\end{equation}

We assume isotropic emission in the shock rest frame. Only when we transform
the radiation emitted into the observer frame, the photons are concentrated
in the forward direction, lying within a cone of half-angle $1/\gamma_{\rm sh}$.
For the afterglow, however, the solid angle is determined by the actual
opening angle of the jet.

At the critical frequency, $\nu_{\rm m}$, in our model the synchrotron cooling 
time is less than the dynamical time, $z_{\rm j}/(\gamma_{\rm sh} c)$. In fact,
in the shock frame

\begin{equation}
t_{\rm s}^{\rm{(sf)}}/t_{\rm d}^{\rm {(sf)}} \approx 7.88 \times 10^{-3} 
(\dot M_{-5 \rm j}^{-1} v_{0.3}) (\epsilon \theta_{-1 \rm j}^{2} 
\gamma_{{\rm m},2}^{-2}) t_5 
\end{equation}
so the synchrotron time scale is shorter than the dynamical time 
scale for our standard parameters.

Considering the cooling evolution of our electron power law spectrum,
to calculate the flux we use the formula $F_{\rm {\nu}}^{\rm {(ob)}}(t) = {dP
\overwithdelims () {d \nu}}^{\rm {(ob)}}_{{\rm p}=2} z^2 x_{\rm {\nu}} \pi 
\theta^2_{\rm j} {1 \over {4 \pi D^2}}$, where ${dP \overwithdelims () {d 
\nu}}^{\rm {(ob)}}_{{\rm p}=2}$ is the total emission power per unit volume 
per unit frequency for an electron power law distribution with an index $p=2$, 
(e.g.~Sect. 6.4 of Rybicki $\&$ Lightman, 1979). In our model this corresponds 
to a flux in the observer frame given by

\begin{eqnarray}
F_{\rm {\nu}}^{\rm {(ob)}}(t) 
& = & 4.00 \times 10^{-18} C_{\rm e}^{\rm {(sf)}}(t) 
      B^{3/2 \rm{(sf)}}(t) \nu^{-1/2} z^2(t) \times \nonumber \\
&   & x_{\rm {\nu}} \theta^2_{\rm j} \gamma_{\rm sh}^3 \times {1 \over {4 D^2}}.
\end{eqnarray}

Here, the quantity $x_{\rm {\nu}}$ represents the thickness of the radiating
shock front, extending behind the shock front to a point where the
local $\nu_{\rm m}$ drops below $\nu$

\begin{eqnarray}
x_{\rm {\nu}} 
& = & 4.70 \times 10^{13} (E_{51}^{3/8} \dot M_{-5 \rm j}^{-9/8} 
      v_{0.3}^{9/8}) \times \nonumber \\
&   & (\epsilon^{3/4} \theta_{-1 \rm j}^{3/2} \gamma_{{\rm m},2}^{-3/4})
      t_5^{9/8} \nu_{14}^{-1/2} \quad \rm{cm}
\end{eqnarray}
where $\nu_{14} = \nu/(10^{14} \rm{Hz})$. In both equations (11) and
(12) $\nu$ is in the shock frame.

Substituting the values of the equations (4), (5), (7), (8) and (12) in
(11), and with $\nu$ in the observer frame

\begin{eqnarray}
F_{\rm {\nu}}^{\rm{(ob)}}(t) 
& = & 7.45 \times 10^{-28} \delta (E_{51}^{5/4} \dot
      M_{-5 \rm j}^{-1/4} v_{0.3}^{1/4}) \times \nonumber \\
&   & \gamma_{{\rm m},2} D_{28.5}^{-2} t_5^{-5/4} \nu_{14}^{-1}
      \quad \rm{erg \enskip cm^{-2} s^{-1} Hz^{-1}}
\end{eqnarray}
where D is the luminosity distance, $D_{28.5} = D/(10^{28.5} {\rm{cm}})$ 
corresponds to a redshift of about $z \simeq 1.5$, using $q_{\rm o} = 1/2$ and 
$H_{\rm o} = 65 \enskip \rm{km \enskip s^{-1} Mpc^{-1}}$.

The flux at frequencies $\nu$ below the critical frequency, $\nu_{\rm m}$,
can be obtained using the flux $F_{{\rm \nu}_{\rm m}}(t)$ at frequencies $\nu >
\nu_{\rm m}$, calculated at the frequency $\nu_{\rm m}$, that is 
$F_{{\rm \nu} < {\rm \nu}_{\rm m}} = F_{{\rm \nu}_{\rm m}}(t) \times 
{\nu \overwithdelims () \nu_{\rm m}}^{\rm{\beta}}$. The corresponding flux 
is given by

\begin{eqnarray}
F_{\nu < \nu_{\rm m}}^{\rm{(ob)}} (t) 
& = (2.84 \times 10^{-28}) {(2.61)}^{- \rm{\beta}} \delta (E_{51}^{3/4 - 
  \rm{\beta}/2} \dot M_{-5 \rm j}^{-1/4} \times \nonumber \\
& v_{0.3}^{1/4}) (\epsilon ^{1/2 + \rm{\beta}/2} \theta_{-1 \rm j}^{1 + 
  \rm{\beta}} \gamma_{{\rm m},2}^{-3/2 -5 \rm{\beta}/2}) D_{28.5}^{-2} 
  \times \nonumber \\ 
& t_5^{1/4 + 3 \rm{\beta}/2} \nu_{14}^{\rm{{\beta}}} {(1+z)}^{1+ \rm{\beta}} 
  \, \rm{erg \enskip cm^{-2} s^{-1} Hz^{-1}}
\end{eqnarray}
For optically thin emission, $\beta$ is equal 1/3 and the time dependence
is $t^{3/4}$. This suggests a gentle optical rise.

Another important check is to calculate the synchrotron self-absorption
coefficient $\alpha_{\rm {\nu}}$ in our model (Sect. 6.8 of Rybicki $\&$ 
Lightman, 1979). Integrating it over an isotropic distribution of particles 
and using the value obtained for the coefficient $C_{\rm e}$, we find 
$\alpha_{\rm {\nu}}(t)$. The optical depth along the path of a travelling 
ray is equal to the product of the absorption coefficient $\alpha_{\rm 
{\nu}}(t)$ times the width of the shell (equation 12), and it will be equal 
to unity at the time

\begin{eqnarray}
t_{{\rm \tau}=1} 
& = & 0.12 \times \delta^{8/15} (E_{51}^{-1/3}
      \dot M_{-5 \rm j} v_{0.3}^{-1}) (\epsilon^{-2/15}
      \theta_{-1 \rm j}^{-4/3} \times \nonumber \\
&   & \gamma_{{\rm m},2}^{2/3}) {\nu}_{14}^{-28/15} \qquad{\rm s}
\end{eqnarray}

\section{Discussion and conclusion}

We now check our model against the observed data. Here we assume a fixed
opening jet angle and a minimum electron Lorentz factor, and the equipartition 
parameters $\epsilon$ and $\delta$ are referenced to unity.

For GRB970208, using the equation (9), at a time of $t_{\rm {opt}}= t/(7.6 
\times 10^4 s)$, we obtain a frequency $\nu_{\rm m}(t_{\rm {opt}}) \approx 
(3.95 \times 10^{14}) {(1+z)}^{-1} E_{51}^{1/2} (\epsilon^{-1/2} \theta_{-1 
\rm j}^{-1} \gamma_{{\rm m},2}^{5/2}) t_{\rm {opt}}^{-3/2}$ Hz.

Substituting the same numbers into the equations for the flux (13), we have 
$F_{\rm {\nu}}(t_{\rm {opt}}) \approx 3.09 \times 10^{-28}$ $\delta 
(E_{51}^{5/4} \dot M_{-5 \rm j}^{-1/4}$ $v_{0.3}^{1/4}) \gamma_{{\rm m},2}$ 
$D_{28.5}^{-2}$ $\nu_{\rm {opt}}^{- 1}$ $t_{\rm {opt}}^{-5/4}$ ${\rm erg 
\enskip cm^{-2} s^{-1}}$ $\rm{Hz^{-1}}$, where $\nu_{\rm {opt}} = \nu/(3.4 
\times 10^{14} \rm{Hz})$.

Again using GRB970208 as reference, we obtain for the flux in the X-band at a 
frequency of $10^{18}$ Hz and a time $3.6 \times 10^4$ s: 
$F_{\rm {\nu}}(t_{\rm X}) \approx 2.67 \times 10^{-31} \delta (E_{51}^{5/4}$ 
$\dot M_{-5 \rm j}^{-1/4} v_{0.3}^{1/4}) \gamma_{{\rm m},2} D_{28.5}^{-2}$ 
$\nu_{\rm {X}}^{- 1} t_{\rm {X}}^{-5/4}$ $\rm{erg \enskip cm^{-2} s^{-1} 
Hz^{-1}}$. 

For the optical depth in the optical band we obtain $\tau_{\rm {\nu}}(t_{\rm 
{opt}}) \approx 2.00 \times 10^{-13} \delta (E_{51}^{-5/8} \dot M_{-5 \rm 
j}^{15/8}$ $v_{0.3}^{-15/8}) (\epsilon^{-1/4} \theta_{-1 \rm j}^{-5/2}$ 
$\gamma_{{\rm m},2}^{5/4})$ ${\nu}_{\rm {opt}}^{-7/2} t_{\rm {opt}}^{-15/8}$, 
hence, within our model the optical emission region is optically thin, except 
very early.

In our model, the condition $1/\gamma_{\rm sh} < \theta_{\rm j}$ corresponds 
to the early phase, when $\nu_{\rm m}$ is actually in the gamma-ray regime. In 
this phase, the corresponding maximum flux in the observer frame is $F_{\rm 
max} \approx 8.66 \times 10^{-7} \delta E_{51} D_{28.5}^{-2} \theta_{-1 \rm 
j}^{-2} \gamma_{{\rm m},2} t^{-1}$ $\rm{erg \enskip cm^{-2} s^{-1}}$. This is 
derived from using all available energy in energetic electrons, and 
redistributing it into emission; photons are generated by various emission 
processes, including pion decay, upscattered by inverse Compton emission, and 
redistributed in photon energy by pair opacity (Rachen \& M{\'e}sz{\'a}ros 
1998). The integration of this flux in time from $10^{-3}$ s to 10 s gives a 
value of about $10^{-5} \delta E_{51} D_{28.5}^{-2} \theta_{-1 \rm j}^{-2} 
\gamma_{{\rm m},2}$ $\rm{erg \enskip cm^{-2}}$. This we identify with the 
initial gamma-ray emission.  

Concerning the variability at times when ${1/ \gamma_{\rm sh} > \theta_{\rm 
j}}$, we note that in observed jets, one often finds inhomogeneities on the 
scale of a few jet-diameters parallel to the axis, and down to some 
fraction of the diameter perpendicular to the axis. However, any 
variability derived from such inhomogeneities is smeared out by 
arrival time differences for the observer.  Therefore, using the jet 
diameter as a reference scale, this smearing limits any temporal 
variability 

\begin{equation}
{\Delta t \over t} \simge {(\gamma_{\rm sh} \theta_{\rm j})}^2 \simeq 0.47 
(E_{51}^{1/2} \dot M_{-5 \rm j}^{-1/2} v_{0.3}^{1/2}) t_5^{-1/2} 
\end{equation}

Inhomogeneities on transverse scales smaller than the jet diameter can
shorten this. This variability may explain the complex features of the 
optical rise in GRB970508 (Galama et al. 1998).

The way in which our model can reproduce the observed properties of GRBs 
depends strictly on two sets of parameters, one set which is verifiable, 
because it derives from known active binary stars, and a second set which 
characterizes the explosion. The explosion depends on the initial shock energy 
$E_{51}$, and the equipartition parameter $\delta$. On the other hand, the 
binary set is composed of the mass loss rate $\dot M_{-5 \rm j}$ of the jet, 
the jet opening angle $\theta_{\rm j}$, the bulk velocity in the jet $v_{0.3}$ 
and the equipartition parameter $\epsilon$. To produce shock waves and the 
observed emission in the jet, it is necessary that the initial amount of matter 
in the shock is comparable with the mass in the jet. This implies that the 
binary system necessary in our model has to have a super Eddington accretion 
rate to get the required mass loss rate. In our Galaxy, one such binary system 
is known, SS433 (Murata and Shibazaki, 1996), for which $\dot M_{\rm disk} 
\simeq (2 \times 10^{-4} M_{\rm {\odot}}/\rm{yr})$. In our model, we have used 
the parameters characteristic of this system, implying that there are system 
like SS433 where the central object indeed is a neutron star.

In the context of the Falcke $\&$ Biermann jet-disk model, SS433 is a
radio-weak jet-disk system, while here we have used their radio-loud
model; but it has been noted that systems such as GRS1915+105 and
perhaps also SS433 can switch from radio-weak to radio-loud in the
terminology of Falcke \& Biermann (1995, 1999). Considering that the
jet ejection rate is bound to the disk accretion rate by the relation 
$\dot M_{\rm jet} \simeq 0.05 \dot M_{\rm disk}$, we find that for an initial 
bulk Lorentz factor $\Gamma_{\rm o} = 10^4$, the mass in the shock $M_{\rm sh} 
\approx 5 \times 10^{-8} M_{\rm {\odot}}$ is equal to the mass present in the 
jet. Therefore, SS433 might be a good candidate to explode as the next 
violent GRB in our Galaxy.

To summarize, our model can explain the initial gamma ray burst, the
spectrum and temporal behaviour of the afterglows, the low baryon load,
an optical rise, and do all this with a modest energy budget. Moreover,
this GRB model is developed within an existing framework for galactic jet
sources, using a set of well determined parameters.

Obviously, we have simplified in many places, using a naive version of 
shock acceleration, only strong shocks, a conical jet geometry, etc., 
but the good agreement with the data we obtain within our framework shows 
that more detailed calculations may be worthwhile.

\vspace{0.3 cm}

\begin{acknowledgements} We thank T.~A.~En{\ss}lin, N.~Masetti, 
P.~M{\'e}sz{\'a}ros, M.~Ostrowski, L.~Piro, and I.~Roussev for useful 
discussions. GP is supported by a DESY grant 05 3BN62A 8. HF is supported 
by a DFG grant 358/1-1\&2.
\end{acknowledgements}


\begin{thebibliography}{99}

\bibitem{ein}
Biermann P.L., 1993, A\&A 271, 649 (section 9)

\bibitem{swei}
Bisnovatyi-Kogan G.S., 1970, Sov.Astr. A.J. 14, 652


\bibitem{uno}
Dar A., 1998, ApJ Lett. 500, L93

\bibitem{sdc}
De Hoffmann F. and Teller E., 1950, Phys. Rev., 80, 682

\bibitem{undi}
Falcke H., Biermann P.L., 1995, A\&A 293, 665

\bibitem{fier}
Falcke H., Biermann P.L., 1999, A\&A 342, 49

\bibitem{qua}
Frail D.A. et al., 1997, Nature 389, 261

\bibitem{qui}
Galama T.J. et al., 1998, ApJ Lett. 497, L13

\bibitem{sei}
Gorosabel J. et al., 1998, A\&A Lett. 335, L5

\bibitem{due}
Guarnieri A. et al., 1997, A\&A Lett. 328, L13

\bibitem{funf}
Kardashev N.S., 1970, Sov.Astr.A.J. 14, 375

\bibitem{cin}
Kulkarni S.R. et al., 1998, Nature 393, 35

\bibitem{sechs}
LeBlanc J.M., Wilson J.R., 1970, ApJ 161, 541

\bibitem{dact}
Marscher A.P., Gear W.K., 1985, ApJ 298, 114 

\bibitem{set}
M{\'e}sz{\'a}ros P., Rees M.J., 1993, ApJ. 405, 278

\bibitem{otto}
M{\'e}sz{\'a}ros P., Proc. 17th Texas Conf. Relativistic Astrophysics,
NY Acad.Sci., 1994

\bibitem{tre}
Metzger M.R. et al., 1997, Nature 387, 878

\bibitem{dict}
Murata K., Shibazaki N., 1996, PASP 48, 819

\bibitem{tredic}
Piro L. et al., 1998, A\&A 329, 906

\bibitem{die}
Paczy{\'n}ski B., 1986, ApJ Lett. 308, L43

\bibitem{und}
Paczy{\'n}ski B., Rhoads J.E., 1993, ApJ Lett. 418, L5

\bibitem{nove}
Panaitescu A., M{\'e}sz{\'a}ros P., 1998, ApJ 492, 683

\bibitem{598} Rachen J.P. \& M\'{e}sz\'{a}ros P., 1998,
Phys. Rev., D58, 123005 

\bibitem{can}
Romanova, M.M., Lovelace, R.V.E., 1997, ApJ 475, 97

\bibitem{cen}
Romanova, M.M. et al., 1997, ApJ 482, 708

\bibitem{two}
Rybicki G.B., Lightman A.P., 1979, ``Radiative Processes in
Astrophysics'', John Wiley \& Sons, New York

\bibitem{dod}
Sari R., Piran T., 1997, ApJ 485, 270

\bibitem{qnd}
Vietri M., 1997, ApJ Lett. 488, L105

\bibitem{tdc}
Waxman E., 1997a, ApJ Lett. 485, L5

\bibitem{qtr}
Waxman E., 1997b, ApJ Lett. 491, L19


\end{thebibliography}
\end{document}